**Footnotes**

1. Stirling's Approximation: $\Gamma(x + 1) = \sqrt{2\pi x}\ x^x e^{-x}\left[1 + \dfrac{1}{12x} + \dfrac{1}{288x^2} + o(\dfrac{1}{x^2})\right]$. The higher order terms may be found in [Abramowitz && Stegun].

dominant term of the asymptotic Bayesian test statistic was shown to be a mutual information. The Bayesian method provided a means for comprehending both the historical test statistic and test procedure. The historical method is incomplete because it does not rigorously handle small samples. The Bayesian method explicitly states all assumptions, whereas the historical method makes assumptions implicitly. The Bayesian method indicates the proper response to take when using the historical method and symbol counts of zero occur, and it provides a justification for placing the cutoff in the historical test only on the high end. In retrospect, it appears that the historical procedure relies heavily upon intuition to provide a useful test statistic and test procedure, whereas the Bayesian method relies heavily on the subjective quantification of prior knowledge.

**Acknowledgments.**

I wish to thank those who have encouraged me and taught me so much about the Bayesian and Non-Bayesian statistical arts over the years, most especially my primary mentor David Wolpert, and Kenneth Hanson. I wish to thank Tom Loredo for reading and correcting an early draft of this paper. This work was done at Los Alamos with the support of the United States department of Energy under contract number W-7405-ENG-36.

When any of these assumptions are violated, we are forced to apply the Bayesian test to gain any understanding whatever of the relative posterior probabilities of hypotheses. In fact, should we desire information of the sort that Eqs. (10) and (11) supply, only the Bayesian procedure is relevant.

The notion of optimality is clear for the Bayesian test. However, for the historical test the notion of Neyman-Pearson (NP) optimality is meaningless. To see this, recall that for fixed significance a NP optimal test minimizes the supremum of the sampling probability that the independent hypothesis is chosen given that the dependent hypothesis is true, where the supremum is taken as the underlying pf in the dependent hypothesis ranges over the dependent hypothesis space. Because it is always possible to choose a dependent hypothesis as close to the independent hypothesis space as may be desired, any dependent hypothesis sampling distribution continuous in the underlying pf's may be brought as close as desired to any independent hypothesis sampling distribution. Thus, for all tests, the supremum indicated above is the same. Thus, no NP optimal test exists for the historical method.

The Bayesian approach clarifies a confusion associated with the historical test. This confusion concerns whether the cutoff for the historical test should be on high values of the test statistic alone, or whether both a high cutoff and a low cutoff are reasonable [2]. Recall from the discussion of Sec. 1 that, in the context of the historical test procedure, the significance level sets a condition only on the probability of rejection of independence, given independence. The region of the sampling distribution chosen for acceptance is therefore arbitrary as long as it excludes a rejection region, in probability, of the given significance. The Bayesian approach strongly argues that, for the particular hypothesis testing problem considered here, the acceptance cutoff should *only* be on the high end.

Finally, note that there is a choice that must be made when the space of symbols being sampled is continuous, that choice being the grouping into discrete symbols the continuous values that may appear. Clearly this choice is crucial for making either testing procedure give useful results. For both procedures there is no definitive method for making this binning choice. However, for the Bayesian procedure there is at least the certainty that the assumptions involved in the binning choice can be made explicit. A prior for the binning choicesmay be assigned, and all calculations for the joint distribution of one or more of hypothesis, pf, and binning choice may be made exactly.

**7.0 Conclusion.**

The Bayesian and historical methods for hypothesis testing have been presented and discussed. The results for the Bayesian method are exact and were found in a direct manner. The historical method uses several quantities of tenuous rigor, and these have been understood by relating them to the exact results of the Bayesian method in the asymptotic limit of large sample size. The The approximation (Eqs. (14a)-(14d))) holds when all $n_{ij}$, $n_{i\cdot}$, or $n_{\cdot j}$ are nonzero. When any of these are zero, simply remove them from the appropriate summations whenever they occur, because $\text{Log}(\Gamma(1)) = 0$ (see Eq. 13). Note that the highest order term is the mutual information between the estimated pf's $n_{ij}/N$ and $(n_{i\cdot}/N)(n_{\cdot j}/N)$. Letting the true underlying pf be $\mathbf{p}^0$, and assuming it factors so that $p_{ij}^0 = p_{i\cdot}^0 p_{\cdot j}^0$, it is easily shown that in the asymptotic regime this term is closely related to Pearson's chi-squared statistic. Let $\Delta_{ij} \equiv n_{ij}/N - p_{ij}^0$, with similar definitions for $\Delta_{i\cdot}$ and $\Delta_{\cdot j}$. In the asymptotic regime and when $\Delta_{ij} \ll p_{ij}^0$ (as would be typical in large samples), the highest order term in the approximation (Eq. 14a) is

$$N\Sigma_{i,j=1}^{r,s} (n_{ij}/N) \text{Log}\left(\frac{(n_{ij}/N)}{(n_{i\cdot}/N)(n_{\cdot j}/N)}\right) \sim \frac{N}{2}\left(\Sigma_{i,j=1}^{r,s}\frac{\Delta_{ij}^2}{p_{i\cdot}^0 p_{\cdot j}^0} - \Sigma_{i=1}^{r}\frac{\Delta_{i\cdot}^2}{p_{i\cdot}^0} - \Sigma_{j=1}^{s}\frac{\Delta_{\cdot j}^2}{p_{\cdot j}^0}\right)$$
$$+ O(N\Delta^3) \quad . \quad (15)$$

We note that $N\Sigma_{i,j=1}^{r,s}\Delta_{ij}^2/(p_{i\cdot}^0 p_{\cdot j}^0)$ is the Pearson chi-squared statistic, distributed as $\chi_{rs-1}^2$. The other terms are chi-squared statistics distributed as $\chi_{r-1}^2$ and $\chi_{s-1}^2$ respectively, and a constant asymptotically proportional to the true mutual information. Because a $\chi_n^2$ distribution has mean n and variance 2n [7], then, for sufficiently large values of r and s, it is possible to conclude that the dominant term in Eq. (15) is the $\chi_{rs-1}^2$ distributed term, $\Sigma_{i,j=1}^{r,s}\Delta_{ij}^2/p_{ij}^0$.

### 6.0 Comparisons of the tests.

The calculations of the last section allow several definitive comparisons of the Bayesian and historical tests to be made. We begin by interpreting the quantities that appear in the historic test in terms of Bayesian test quantities.

In the comments at the end of the last section we noted that the dominant term in the asymptotic form of $\text{Log}(CR(\mathbf{n}))$ arising in the Bayesian hypothesis test is predominantly equal to one half of the Pearson chi-squared statistic of the historical test, given that the dimensions r and s are sufficiently large and that *the true hypothesis is a member of the independent set*. Insofar as the Pearson chi-squared statistic is asymptotically equal to the historical test statistic, we may conclude that the historical test is loosely based on *twice the logarithm of the minimum risk (error) Bayesian test*. The significance $\alpha$ is therefore loosely the probability that twice the logarithm of the Bayesian test statistic will exceed the cutoff value c of the historic test. Since the Bayesian test statistic is the ratio of the posterior probabilities of the dependent and independent hypotheses, we may conclude that the historical test is based on the same, in the asymptotic regime, and for sufficiently large dimensions r and s.

In the small sample regime or when the hypothesis is not necessarily a member of the independent set (the typical case, otherwise why bother testing at all?), the discussion of the last paragraph does not hold. Several constants have been neglected (the ratio of the priors, the dimension dependent term). The previous discussion also breaks down if the dimensions r and s are small.

The **n** independent factor, $\Gamma(rs)/(\Gamma(r)\Gamma(s))$, is dependent only upon the dimensions of the parameter spaces involved. When no observations have been made (N = 0) it is cancelled exactly by the factor $\Gamma(N+r)\Gamma(N+s)/\Gamma(N+rs)$. In the next section we relate Eq. (12) to an estimated mutual information function and the historical chi-squared test. For now, note that if it is desired to minimize the risk of error (making an incorrect hypothesis choice) we would choose I or $\bar{I}$ based on the larger of $P(I \mid \mathbf{n})$ and $P(\bar{I} \mid \mathbf{n})$. This leads to the Bayesian minimal error test procedure of choosing I if $CR(\mathbf{n}) < 1$, choosing $\bar{I}$ if $CR(\mathbf{n}) > 1$, and otherwise choosing randomly (with equal probability).

### 5.0 Asymptotics of the Bayesian test and Mutual Information.

The previous two sections described the historical (Sec. 3) and the Bayesian (Sec. 4) tests for independence. The easily computed statistic $CR(\mathbf{n})$ was identified as the important object in the Bayesian test. In this section we develop the asymptotic form of $CR(\mathbf{n})$, show how the dominant term of this asymptotic form may be viewed as an estimated mutual information, and relate this asymptotic form to the chi-squared form in the historical test, which is by default an asymptotic test.

Taking the logarithm of Eq. (12) gives us the sum of several terms. In finding the asymptotic form of $\text{Log}(CR(\mathbf{n}))$ there are six basic terms which are easily approximated and summed, along with the N-independent terms. The six basic terms are the following:

$$\Sigma_{i,j=1}^{r,s} \text{Log}(\Gamma(n_{ij}+1)), \; -\Sigma_{i=1}^{r} \text{Log}(\Gamma(n_{i\cdot}+1)), \; -\Sigma_{j=1}^{s} \text{Log}(\Gamma(n_{\cdot j}+1)),$$
$$\text{Log}(\Gamma(N+r)), \; \text{Log}(\Gamma(N+s)), \; \text{and} \; -\text{Log}(\Gamma(N+rs)). \tag{13}$$

Using Stirling's formula[1] and carrying out the approximations to $o(1/N)$ shows that $CR(\mathbf{n})$ is the sum of terms of various orders in N. We write these terms in decreasing order:

Order N: $$N\Sigma_{i,j=1}^{r,s} (n_{ij}/N) \text{Log}\left(\frac{(n_{ij}/N)}{(n_{i\cdot}/N)(n_{\cdot j}/N)}\right) \tag{14a}$$

Order Log(N):
$$\Sigma_{i,j=1}^{r,s} \text{Log}(n_{ij}) - \Sigma_{i=1}^{r} \text{Log}(n_{i\cdot}) - \Sigma_{j=1}^{s} \text{Log}(n_{\cdot j}) - (rs-r-s+1/2)\text{Log}(N). \tag{14b}$$

Order 1: $$\left(\frac{rs-r-s+1}{2}\right) \text{Log}(2\pi) + \text{Log}\left(\frac{\Gamma(rs)}{\Gamma(r)\Gamma(s)}\right) + \text{Log}(C) \tag{14c}$$

Order 1/N:
$$\frac{1}{12N}\left(\Sigma_{i,j=1}^{r,s} n_{ij}^{-1} - \Sigma_{i=1}^{r} n_{i\cdot}^{-1} - \Sigma_{j=1}^{s} n_{\cdot j}^{-1} + 1\right) - \frac{1}{2N}\left((r^2s^2-rs) + (r^2-r) + (s^2-s)\right) \tag{14d}$$

For the independent case in Eq. (3) we may rewrite the integral for $P(\mathbf{n} \mid I)$ as

$$P(\mathbf{n} \mid I) \propto \binom{N}{\mathbf{n}} \left( \int \Pi_{i=1}^{r} (p_i^a)^{n_i\cdot} \delta(\Sigma_{i=1}^{r} p_i^a - 1) d\mathbf{p}^a \right) \times \left( \int \Pi_{j=1}^{s} (p_j^b)^{n_{\cdot j}} \delta(\Sigma_{j=1}^{s} p_j^b - 1) d\mathbf{p}^b \right). \quad (5)$$

Similarly, for the dependent case in Eq. (3) we may write the integral for $P(\mathbf{n} \mid \bar{I})$ as

$$P(\mathbf{n} \mid \bar{I}) \propto \binom{N}{\mathbf{n}} \int \Pi_{i,j=1}^{r,s} p_{ij}^{n_{ij}} \delta(\Sigma_{i,j=1}^{r,s} p_{ij} - 1) d\mathbf{p}. \quad (6)$$

In Eqs. (5) and (6) note that, because of normalization of the priors, the $N = 0$ values of the right hand sides of the proportionalities Eqs. (5) and (6) should be 1. The integrals in Eqs. (5) and (6) are found using convolution and Laplace transform techniques described in [5]. The integration shows that the normalization constant for the independent prior is $\Gamma(r)\Gamma(s)$, while that of the dependent prior is $\Gamma(rs)$. Taking into account the normalization constants for the priors, the results for the hypothesis-conditioned sampling distributions are

$$P(\mathbf{n} \mid I) = \Gamma(r)\Gamma(s) \binom{N}{\mathbf{n}} \frac{\Pi_{i=1}^{r} \Gamma(n_{i\cdot} + 1) \Pi_{j=1}^{s} \Gamma(n_{\cdot j} + 1)}{\Gamma(N+r)\Gamma(N+s)}, \quad (7a)$$

and

$$P(\mathbf{n} \mid \bar{I}) = \Gamma(rs) \binom{N}{\mathbf{n}} \frac{\Pi_{i,j=1}^{r,s} \Gamma(n_{ij} + 1)}{\Gamma(N+rs)}. \quad (7b)$$

Let $C \equiv P(\bar{I})/P(I)$ and $R(\mathbf{n}) \equiv P(\mathbf{n} \mid \bar{I})/P(\mathbf{n} \mid I)$. Using Eq. (2), the posterior probabilities of the hypotheses may be rewritten in terms of known quantities as

$$P(I \mid \mathbf{n}) = 1/(1 + CR(\mathbf{n})). \quad (8)$$

Similarly,

$$P(\bar{I} \mid \mathbf{n}) = CR(\mathbf{n})/(1 + CR(\mathbf{n})). \quad (9)$$

It should be pointed out that the probabilities for the hypotheses given the data in Eqs. (8) and (9) are exact and hold for all N.

Other desired quantities are the posterior probabilities that a pf occurs given the observed data and hypothesis, $P(\mathbf{p} \mid \mathbf{n}, I)$, and $P(\mathbf{p} \mid \mathbf{n}, \bar{I})$. Using Bayes Theorem, write these in terms of known quantities (prior equation and Eqs. (4), (5)) as

$$P(\mathbf{p} \mid \mathbf{n}, I) = P(\mathbf{n} \mid \mathbf{p}, I) P(\mathbf{p} \mid I)/P(\mathbf{n} \mid I), \quad (10a)$$

and similarly (prior equation and Eqs. (4), (6))

$$P(\mathbf{p} \mid \mathbf{n}, \bar{I}) = P(\mathbf{n} \mid \mathbf{p}, \bar{I}) P(\mathbf{p} \mid \bar{I})/P(\mathbf{n} \mid \bar{I}). \quad (10b)$$

Another quantity of interest is the probability of any pf given data only, which is written in terms of known quantities (Eqs. (8), (9), and (10)) as,

$$P(\mathbf{p} \mid \mathbf{n}) = P(\mathbf{p} \mid \mathbf{n}, I) P(I \mid \mathbf{n}) + P(\mathbf{p} \mid \mathbf{n}, \bar{I}) P(\bar{I} \mid \mathbf{n}). \quad (11)$$

At this point it is of interest to examine the function $R(\mathbf{n})$. Explicitly,

$$R(\mathbf{n}) = \frac{\Pi_{i,j=1}^{r,s} \Gamma(n_{ij} + 1)}{\Pi_{i=1}^{r} \Gamma(n_{i\cdot} + 1) \Pi_{j=1}^{s} \Gamma(n_{\cdot j} + 1)} \times \frac{\Gamma(N+r)\Gamma(N+s)}{\Gamma(N+rs)} \times \frac{\Gamma(rs)}{\Gamma(r)\Gamma(s)}. \quad (12)$$

Note that $R(\mathbf{n})$ consists of a factor that is dependent upon $\mathbf{n}$ and a factor that is independent of $\mathbf{n}$.

that the pf factors as in Sec. 2 (independence), and the second that it does not (dependence). Clearly, the independent hypothesis has lower dimensionality than the dependent hypothesis. Indeed, much as we would separate a line from a plane, we may look at the choice of the two hypotheses as having divided a higher dimensional space, the space of all pf's, into a lower dimensional subspace and a space with the same dimensionality as the original. Let the prior probability of the independent hypothesis, I, be $P(I)$, and similarly, let the prior probability of the dependent hypothesis, $\bar{I}$, be $P(\bar{I})$. Clearly $P(I) + P(\bar{I}) = 1$. Further, let the prior on each hypothesis space of pf's be *uniform*, or constant, so that $P(\mathbf{p} \mid I) = c_I$ and $P(\mathbf{p} \mid \bar{I}) = c_{\bar{I}}$. The constants are such that each is normalized. Note that each prior is zero for pf's outside the respective hypothesis spaces. Because both of the hypothesis spaces contain pf's that are *constrained* by having components summing to one it is important to explicitly state the form of the priors $P(\mathbf{p} \mid I)$ and $P(\mathbf{p} \mid \bar{I})$. When the pf factors, the rs probabilities in each pf are specified by $r + s$ underlying parameters, which in this problem represent marginal probabilities. The constraint $\sum_{i,j=1}^{r,s} p_{ij} = 1$ reduces the number of independent parameters to $r + s - 2$. When the pf does not factor, there are rs parameters, with the constraint reducing the number of independent parameters to $rs - 1$. For the independent case, we choose as the underlying parameters two pf's, $\mathbf{p}^a$ and $\mathbf{p}^b$, each of dimension r and s respectively, and incorporate the constraints that the components of the respective pf's sum to one using delta functions, so that $P(\mathbf{p} \mid I) = P(\mathbf{p}^a, \mathbf{p}^b \mid I) \propto \delta(\sum_{i=1}^{r} p_i^a - 1)\delta(\sum_{j=1}^{s} p_j^b - 1)$. Similarly, for the dependent case we parameterize the space using $\mathbf{p}$ itself, so that $P(\mathbf{p} \mid \bar{I}) \propto \delta(\sum_{i,j=1}^{r,s} p_{ij} - 1)$. It is not difficult to show that, with the priors chosen in this manner, the densities on the corresponding surfaces of constraint are indeed constant [4]. The assumption of uniformity is driven by the subjective desire to put as little knowledge as possible into the prior over the pf's, and to provide for simplicity of calculation. Uniformity is one requirement that we may drop: when the prior is not uniform the method for computing results in a manner similar to those presented in this paper is presented in [6]. Related hypothesis testing work appears in [8].

A desired quantity is the *posterior* probability of independence given data, $P(I \mid \mathbf{n})$. From it, because there are only two mutually exclusive hypotheses, we find that the probability of dependence given data is $P(\bar{I} \mid \mathbf{n}) = 1 - P(I \mid \mathbf{n})$. Using Bayes Theorem the posterior hypothesis probabilities are given by

$$P(I \mid \mathbf{n}) = P(\mathbf{n} \mid I)P(I)/P(\mathbf{n}), \text{ and } P(\bar{I} \mid \mathbf{n}) = P(\mathbf{n} \mid \bar{I})P(\bar{I})/P(\mathbf{n}). \quad (1)$$

where $P(\mathbf{n} \mid I)$ is the likelihood, or sampling distribution, given that hypothesis I is true, and similarly $P(\mathbf{n} \mid \bar{I})$ is the likelihood given that hypothesis $\bar{I}$ is true. Again, because there are only two mutually exclusive hypotheses,

$$P(\mathbf{n}) = P(\mathbf{n} \mid I)P(I) + P(\mathbf{n} \mid \bar{I})P(\bar{I}). \quad (2)$$

Furthermore,

$$P(\mathbf{n} \mid I) = \int P(\mathbf{n} \mid \mathbf{p}, I)P(\mathbf{p} \mid I)d\mathbf{p}, \text{ and similarly } P(\mathbf{n} \mid \bar{I}) = \int P(\mathbf{n} \mid \mathbf{p}, \bar{I})P(\mathbf{p} \mid \bar{I})d\mathbf{p}, \quad (3)$$

where $P(\mathbf{n} \mid \mathbf{p}, I)$ and $P(\mathbf{n} \mid \mathbf{p}, \bar{I})$ are the likelihoods given fixed pf's in the respective hypotheses. The fixed pf likelihoods in Eq. (3) are multinomial distributions, with

$$P(\mathbf{n} \mid \mathbf{p}, I) = \binom{N}{\mathbf{n}} \Pi_{i,j=1}^{r,s} (p_i^a p_j^b)^{n_{ij}} \text{ and } P(\mathbf{n} \mid \mathbf{p}, \bar{I}) = \binom{N}{\mathbf{n}} \Pi_{i,j=1}^{r,s} p_{ij}^{n_{ij}}. \quad (4)$$

The historical testing approach considered in this section is the chi-squared test for independence. When the underlying pf is represented by a vector **p**, and observed counts are represented by a vector **n**, (both of dimension r), the chi-squared test is based upon the observation that for $N \to \infty$ the distribution of the statistic $Q(\mathbf{n}, \mathbf{p}) = \Sigma_{i=1}^{r} (n_i - Np_i)^2 / Np_i$ converges to a $\chi^2$ distribution with $r - 1$ degrees of freedom (dof) [1] [2] [3]. (Explicitly, the $\chi^2$ probability density function with n dof is given by $\chi_n^2(x) = 2^{-n/2} \Gamma(n/2) x^{n/2-1} e^{-x/2}$, where $x \geq 0$.) (For small N the distribution of Q is somewhat different: we must examine the exact sampling distribution of **n** in order to find the sampling distribution of Q.) In the historical framework the hypotheses being tested are $H_0$: **p** and $H_1$: not **p**. The decision to reject $H_0$ with significance $\alpha \in (0, 1)$ is based on the cumulative distribution function of Q: With c chosen so that $P(Q > c) = \alpha$, $H_0$ is rejected if $Q > c$, otherwise it is accepted. Qualitatively, all is well. The Q functional is proportional to a measure of the 'distance' of **n**/N from **p**. This is easily seen by rearranging Q as $Q(\mathbf{n}, \mathbf{p}) \propto \Sigma_{i=1}^{r} (n_i/N - p_i)^2 / p_i$, where the proportionality constant is N. For $N \to \infty$, the right-hand side of the proportionality converges to zero when the hypothesis is true, otherwise it converges to a nonzero value. With the factor of N present, it either converges to a $\chi_{r-1}^2$ distributed random variable (rv), or it diverges to infinity. The significance is the probability of incorrectly choosing the hypothesis $H_1$ (given $H_0$ is true) for this test. Further, and perhaps most importantly, when **p** is not the true pf, the values of Q are increasing for $N \to \infty$, and the test is based upon rejecting $H_0$ for large Q.

To apply the chi-squared test just outlined to the problem of determination of independence, let the assumed hypotheses be $H_0: p_{ij} = p_{i \cdot} p_{\cdot j}$ and $H_1: p_{ij} \neq p_{i \cdot} p_{\cdot j}$. The quantities $p_{i \cdot}$ and $p_{\cdot j}$ are *estimated from the data* by their maximum likelihood values $\hat{p}_{i \cdot} = n_{i \cdot}/N$ and $\hat{p}_{\cdot j} = n_{\cdot j}/N$. With the estimated marginal distributions found in this manner, the statistic $Q(\mathbf{n}, \mathbf{p}) = \Sigma_{i=1}^{r} \Sigma_{j=1}^{s} (n_{ij} - N\hat{p}_{i \cdot} \hat{p}_{\cdot j})^2 / N\hat{p}_{i \cdot} \hat{p}_{\cdot j}$ is asymptotically ($N \to \infty$) distributed as a $\chi^2$ rv with $(r-1)(s-1)$ dof. [2]. Using this $\chi^2$ distribution, the test for a given significance goes through as before. There are several criticisms that must be made, but we leave these until Sec. 6 where this historical testing method is compared to the Bayesian method.

**4.0 Bayesian test for independence.**

In the Bayesian procedure we are directly interested in, given the observed data, the probabilities of independence and dependence of the underlying pf, as well as any other quantities (averages, uncertainties, etc.) that may be of interest. Later we interpret the procedure in terms of the minimization of a risk. In the process of developing the Bayesian approach to this problem we show that the important underlying quantity of interest is the mutual information function.

The crux of the Bayesian approach is the choice of prior for the two hypotheses, a choice we necessarily make based on *subjective* measures of our *knowledge* of the underlying pf, and with considerations made for assuring the calculability of results under our assumptions. The prior makes explicit all assumptions involved. With this in mind, consider the two hypotheses, the first testing framework and from the Bayesian hypothesis testing framework, assuming an optimality criterion and test, or prior and risk, as appropriate to the method. By comparing the Bayesian hypothesis testing procedure to the historical hypothesis testing procedure, we make clear some of the assumptions that the historical hypothesis testing procedure implicitly makes. For instance, we demonstrate that the notions of optimality in the two frameworks are quite different, with the notion of Neyman-Pearson optimality being vacuous for the problem under consideration. Most importantly, we show that the problem of testing for independence is easily formulated within the Bayesian framework. In the Bayesian framework all assumptions are explicit and there is no need to improvise a testing procedure. Indeed, all quantities of interest are calculable within the Bayesian framework, while it is not clear what quantities are calculable in the historical framework. Also, the Bayesian approach immediately gives a result useful for all sample sizes, whereas the historical procedure does not apply when the sample size is small. Clear benefits of the historical procedure are 1) no need to consider hypothesis probabilities. (In many cases this is not just of theoretical importance the dimensionality of the spaces involved make it a necessity, but this is not the case here.) Clear benefits of the Bayesian method are the ability to 1) quantify all assumptions, 2) understand sensitivity to the assumptions, 3) rigorously determine probabilities of hypotheses given data and assumptions, and 4) rigorously understand the uncertainties involved. We also note that with the Bayesian method there is 5) avoid discovery and analysis of an appropriate testing procedure.

## 2.0 Definition of the problem.

The problem we are considering is that of determining whether the joint *probability function* (pf) $P(X, Y)$ is *independent* given a sample of N observations (ordered pairs of values $(X, Y)$) drawn from this pf. A joint pf is independent when the probability of seeing a certain value of X is independent of the value of Y (or similarly with $Y \leftrightarrow X$). This definition is easily shown to be equivalent to the pf being factorable as $P(X, Y) = P(X)P(Y)$, where $P(X=x_i) = \Sigma_{j=1}^{s} P(x_i, y_j)$ and similarly $P(Y=y_j) = \Sigma_{i=1}^{r} P(x_i, y_j)$. (As indicated, the indices i, j are assumed to run from 1 to integers r and s respectively. That is, there are r distinct possible values (*symbols*) of X and similarly s distinct values of Y.) If a joint pf is not independent then it is *dependent*. For brevity we use the notation $p_{ij} = P(X=x_i, Y=y_j)$, $p_{i \cdot} \equiv P(X=x_i)$ and $p_{\cdot j} \equiv P(Y=y_j)$. Since we will be considering observed data, the notation $n_{ij}$ indicates the number of observations with $X = x_i, Y = y_j$, and similarly $n_{i \cdot} \equiv \Sigma_{j=1}^{s} n_{ij}$ and $n_{\cdot j} \equiv \Sigma_{i=1}^{r} n_{ij}$. The constraints $\Sigma_{i,j=1}^{r,s} n_{ij} = N$ and $\Sigma_{i,j=1}^{r,s} p_{ij} = 1$ are immediate and are assumed to hold through all that follows. At times we consider vectors representing pf's and the corresponding observed data and denote their components $p_i$ and $n_i$ respectively. Whenever possible, and depending upon the context, we represent vectors or matrices in bold type, e.g. **p** = $(p_i)$ or **p** = $(p_{ij})$.

## 3.0 Historical test for independence using chi-squared.

procedure, given data and several hypotheses to be tested for, tests each hypothesis, chooses the hypothesis by examining the test values, and provides a measure of the quality of the test by providing the level of significance of each hypothesis. Further development of the hypothesis testing framework considers the notion of *optimality* of a test. For the two hypothesis case, the *Neyman-Pearson optimal test* with specified significance level is any test that minimizes the probability that the hypothesis is accepted given that it is not true, and has a significance level not surpassing the specified significance level. It is not difficult to show that all Neyman-Pearson optimal tests for simple hypotheses are of the form of a direct comparison of the ratio of the likelihoods for each of the hypotheses to a fixed value.

The historical hypothesis testing procedure ignores the probabilities of occurrence for each of the hypotheses and of the various parameter vectors of the hypotheses. The procedure implicitly assumes that each hypothesis occurs with some fixed probability and that the parameter vectors for each hypothesis occur with some fixed probability, but does not make it clear what these assumptions are. The important indicator that the probabilities of occurrence for each of the hypotheses are ignored is that everything is based on the sampling distribution, or likelihood.

Another framework exists for hypothesis testing that does not make implicit assumptions about the probabilities of occurrence of hypotheses and parameter vectors. In the Bayesian framework [3], a *prior* distribution is chosen that quantifies how the various hypotheses occur. Instead of grouping all parameter vectors in a given hypothesis into a group with unclear probabilities of occurrence, the prior quantifies the probability of occurrence of the parameter vectors. Further, in the Bayesian framework all hypothesis choices are based upon a *risk* function: choose the hypothesis that minimizes the risk. To see why this is important, consider the case where there are two simple hypotheses, but it is known that one is far more probable than the other. Suppose that both hypotheses *explain* the data equally well (where 'explaining equally well' means that both likelihood functions are equal). The historical testing procedure with a strict likelihood-based test will choose one of the hypotheses at random (because they have equal likelihoods), while the Bayes procedure, which considers the probabilities of the hypotheses and minimizes a risk that is the probability of error of choice, will choose the hypothesis that is most probable. In many cases this is the desirable choice. In other cases there may be such a high risk associated with the incorrect choice that the Bayesian method would choose the low-probability hypothesis, even under the conditions outlined above. Another reason it is important to understand the assumptions completely is that often a procedure is desired that can choose between hypotheses parameterized, say, by regions of $R^n$, with a different n for each hypothesis. (A case of this form is being considered in the rest of this paper.) Without clearly specifying the prior, it is difficult to understand how any hypothesis involving a parameter space of lower dimension could be favored over a parameter space of higher dimension, especially if the union of the disjoint spaces of lower dimension and higher dimension covers the space of probability functions.

In what follows we examine a specific hypothesis testing problem - that of testing for independence of an underlying joint probability distribution based on a finite sample of observed data drawn from that distribution. We will consider the problem from both the historical hypothesis

**Mutual Information as a Bayesian Measure of Independence**

David Wolf

Department of Physics,
University of Texas, Austin 78712.

Los Alamos National Laboratory,
Los Alamos, NM 87545.

wolf@lanl.gov

**0.0 Abstract.**

The problem of hypothesis testing is examined from both the historical and the Bayesian points of view in the case that sampling is from an underlying joint probability distribution and the hypotheses tested for are those of independence and dependence of the underlying distribution. Exact results for the Bayesian method are provided. Asymptotic Bayesian results and historical method quantities are compared, and historical method quantities are interpreted in terms of clearly defined Bayesian quantities. The asymptotic Bayesian test relies upon a statistic that is predominantly mutual information.

**1.0 Introduction.**

Problems of hypothesis testing arise ubiquitously in situations where observed data is produced by an unknown process and the question is asked "From what process did this observed data arise?" Historically, the hypothesis testing problem is approached from the point of view of *sampling*, whereby several fixed hypotheses to be tested for are given, and all measures of the *test* and its quality are found directly from the *likelihood*, i.e. by what amounts to sampling the likelihood [2] [3]. (To be specific, a hypothesis is a set of possible parameter vectors, each parameter vector completely specifying a sampling distribution. A *simple hypothesis* is a hypothesis set that contains one parameter vector. A *composite* hypothesis occurs when the (nonempty) hypothesis set is not a single parameter vector.) Generally, the test procedure chooses as true the hypothesis that gives the largest test value, although the notion of procedure is not specific and may refer to any method for choosing the hypothesis given the test values. Since it is of interest to quantify the quality of the test, a *level of significance* is generated, this level being the probability that, under the chosen hypothesis and test procedure, an incorrect hypothesis choice is made. The significance is generated using the sampling distribution, or likelihood. For simple hypotheses the level of significance is found using the single parameter value of the hypothesis. When a test is applied in the case of a composite hypothesis, a *size* for the test is found that is given by the supremum probability (ranging over the parameter vectors in the hypothesis set) that under the chosen hypothesis an incorrect hypothesis choice is made. To summarize, the historical hypothesis testing